# Surface Hydrogen Coverage on Pt/Graphene Measured by Carbon Ion ERDA

Chien-Hsu Chen[1*], Huan Niu[1], Hung-Kai Yu[1], Tsung Te Lin[2] and Yao-Tung Hsu[2]

[1]Accelerator Lab., Nuclear Science and Technology Development Center, National Tsing Hua University, Hsinchu, 300044 Taiwan, ROC

[2]Department of Mechanical and Systems Engineering, National Atomic Research Institute, Taoyuan City, 325207, Taiwan, ROC

achchen@mx.nthu.edu.tw*

Graphene, a two-dimensional monolayer of sp²-bonded carbon atoms in a honeycomb lattice, possesses exceptional electronic, mechanical, and quantum properties, making it highly attractive for energy storage, spintronics, and microelectronics. Functionalizing graphene with platinum (Pt) adatoms can further enhance its properties, particularly for hydrogen storage applications. In this study, we experimentally investigate hydrogen adsorption on Pt-decorated graphene using Elastic Recoil Detection Analysis (ERDA). By irradiating the Pt/graphene film with a 4.1 MeV $C^{2+}$ ion beam and detecting recoiled hydrogen atoms at a 30° scattering angle, we obtain the hydrogen depth profile, providing critical insights into its storage behavior.

Introduction

Hydrogen is widely regarded as a key energy carrier for future low-carbon systems due to its high specific energy and zero-carbon utilization[1, 2]. However, storage and transport of gaseous or liquid hydrogen require either high pressure or cryogenic conditions, which impose significant safety, cost, and energy penalties[3]. This has motivated extensive efforts toward solid-state hydrogen storage materials. Carbon-based materials, including activated carbon, carbon nanotubes, and graphene, have been investigated as potential physisorption media because of their low atomic mass and chemical stability[4, 5]. Nevertheless, hydrogen adsorption in these systems is dominated by weak van der Waals interactions, leading to measurable uptake only under extreme pressure or low-temperature conditions[6]. Reported storage capacities furthermore exhibit poor reproducibility across studies and remain below the targets set by the U.S. Department of Energy[7, 8].

Graphene[2, 9-12], with its single-atom thickness, high surface area, and well-defined two-dimensional structure, has been considered a model platform for hydrogen adsorption studies.

Theoretical calculations predict a maximum uptake of approximately 7.7 wt% for full monolayer hydrogen coverage[6, 13]. However, experimental results reported in the literature vary by orders of magnitude, ranging from negligible adsorption to values exceeding theoretical limits. These discrepancies can be largely attributed to the limitations of indirect measurement methods (gravimetric, volumetric), the difficulty of quantifying hydrogen due to its low mass and high diffusivity, and the absence of analytical techniques capable of directly determining hydrogen content and depth distribution. Consequently, the actual hydrogen uptake capability of graphene remains unresolved[14].

A critical gap in current hydrogen-storage research is the lack of direct and quantitative measurements of hydrogen on graphene. Although several mechanisms have been proposed—including physisorption, chemisorption, and metal-catalyzed spillover—none of these can be reliably validated without accurate hydrogen quantification. Elastic Recoil Detection Analysis (ERDA) is one of the few techniques with the required sensitivity and minimal matrix interference for direct hydrogen analysis[15-17]. However, systematic ERDA measurements on monolayer graphene have rarely been reported.

The objective of this work is to provide quantitative, direct measurements of hydrogen adsorbed on monolayer graphene using the ERDA system available in our laboratory[17]. Hydrogen concentrations in graphene under various processing conditions are measured and compared with theoretical expectations and previous experimental reports. These results establish a reliable baseline for evaluating hydrogen uptake in graphene and clarify long-standing inconsistencies in the literature. Furthermore, the generated ERDA data enable a more accurate assessment of proposed adsorption mechanisms and offer a reference for future material design and characterization in hydrogen-storage applications.

Method

The graphene used in this study was purchased from Graphenea[18]. It consists of a 90 nm $SiO_2$ layer grown on a silicon substrate, with a monolayer of graphene deposited on top. After surface modification, platinum was embedded beneath the graphene, positioned between the graphene layer and the $SiO_2$ layer. The prepared samples were first left in ambient air for a period of time before being transferred into a vacuum chamber. ERDA was performed using the NEC 9SDH-2 accelerator of accelerator laboratory in National Tsing Hua University equipped with a SNICS-II ion source to generate $C^-$ ions, which were accelerated to 72 keV, selected by an analyzing magnet, then further accelerated through the tandem accelerator. A 90° magnet was used to select $C^{2+}$ ions, producing a final energy of 4.1 MeV[19]. At the experimental station, samples were introduced into the vacuum chamber via a load-lock system. The sample was positioned at a 25° angle relative to the incident ion beam, with a PIPS detector placed at a 30° scattering angle as shown in figure 1. SIMNRA[20] simulations indicated that for 4.1 MeV carbon ions, a 6 μm aluminum foil could completely block forward-scattered carbon ions while allowing hydrogen ions to pass through and reach the PIPS detector. This configuration enabled selective measurement of hydrogen ions from elastic collisions, allowing determination of the hydrogen depth profile within the graphene and $SiO_2$/Si layers.

Figure 2 schematically illustrates the experimental sequence for Elastic Recoil Detection (ERD) analysis and hydrogen absorption. (1) First ERD analysis — A carbon ion beam is incident on the sample at an angle of 25° inside a vacuum chamber. Recoiled particles are detected by a PIPS detector positioned at 30° with respect to the beam direction and covered by a 6-μm-thick aluminum foil. (2) Hydrogen absorption — The sample is transferred to a load-lock chamber filled with 14 psi $H_2$ gas for 5 minutes, allowing hydrogen atoms to be absorbed into the sample. (3) Second ERD analysis — After evacuation of the hydrogen gas, the sample is returned to the vacuum chamber for a second ERD measurement to evaluate changes in the hydrogen content. Figure 3 shows the experimental setup for hydrogen gas exposure and subsequent evacuation.

Results

To establish a reliable hydrogen reference, the hydrogen recoil spectrum of a pristine silicon wafer was first measured. The silicon substrate was a (001) n-type wafer with no prior chemical processing, ensuring that it had not been exposed to any hydrogen-containing environment. The resulting spectrum confirms that both the bare graphene sample (without Pt decoration) and the reference silicon wafer exhibit identical hydrogen concentrations. Furthermore, even after the graphene was subjected to carbon-ion irradiation during ERDA measurements, its hydrogen concentration remained indistinguishable from that of both the silicon reference and the undamaged graphene. These results are summarized in Figure 4.

Subsequently, the ion-damaged graphene sample was exposed to hydrogen gas at a pressure of 14 psi for 5 minutes in the load-lock chamber. Despite this controlled hydrogen exposure, the measured hydrogen concentration remained unchanged and consistent with that of the silicon reference. Taken together, these results demonstrate that graphene without Pt decoration does not exhibit measurable hydrogen absorption under the experimental conditions employed in this study.

In contrast, Figure 4 shows that Pt-decorated graphene exhibits markedly different behavior. The ERDA spectra reveal a clear increase in hydrogen concentration on the Pt-decorated surface immediately after exposure to 14 psi $H_2$ gas for 5 minutes, indicating that hydrogen adsorption is significantly enhanced in the presence of Pt. Additional hydrogen exposure cycles in the load-lock chamber lead to a further increase in the hydrogen signal until a saturation level is reached, beyond which no additional hydrogen uptake is observed. This saturation behavior confirms the existence of a finite hydrogen adsorption capacity on the Pt-decorated graphene surface.

SIMNRA simulations indicate that the pristine silicon wafer has a surface hydrogen areal density of approximately 0.3E15 protons/cm², which is consistent with previously reported results by Ecker et al[21]. The simulations also show that the bare graphene sample possesses a similar hydrogen concentration both before and after hydrogen gas exposure. Even after ERDA-induced carbon-ion irradiation, no additional hydrogen uptake is detected on the bare graphene surface. In contrast, the Pt-decorated graphene sample exhibits a pronounced hydrogen signal following hydrogen exposure, with SIMNRA analysis indicating an absorbed hydrogen areal density of approximately 10.3E15 protons/cm².

Discussion

The present results indicate that bare graphene exhibits negligible hydrogen adsorption both before and after ERDA measurements. Figure 5(a) presents the SIMNRA simulation together with the experimental ERDA spectrum for bare graphene exposed to 14 psi $H_2$ gas. The extracted hydrogen areal density is approximately 0.3E15 protons/cm², which is comparable to that obtained for the silicon reference sample. Although carbon-ion irradiation at 4.1 MeV introduces lattice damage into the graphene, this damage alone does not result in a measurable increase in hydrogen uptake[9, 11]. A slight enhancement in the hydrogen signal is observed in the recoil spectra; however, its magnitude is small and remains within experimental uncertainty. Repeated cycles of hydrogen exposure and ERDA analysis yield consistent results, further supporting the conclusion that crystalline damage by itself is not a sufficient condition to enable significant hydrogen adsorption in graphene.

This observation contrasts with some earlier reports suggesting that irradiation-induced defects in graphene can create unsaturated carbon bonds, which may be passivated by hydrogen atoms to form hydrocarbon structures, often referred to as damaged graphene membranes (DGM)[9, 11]. The absence of such behavior in the present study implies that, under the current experimental conditions, defect creation alone does not effectively promote hydrogen trapping, and that additional catalytic or chemical mechanisms are required.

In contrast, Pt-decorated graphene displays a distinctly different hydrogen adsorption mechanism. A pronounced increase in hydrogen concentration is observed immediately after hydrogen gas exposure, indicating that the presence of Pt plays a critical role in facilitating hydrogen uptake. This behavior is consistent with previously proposed hydrogen spillover mechanisms[8, 22], in which Pt catalytically dissociates molecular hydrogen into atomic hydrogen. The resulting hydrogen atoms can migrate across the surface and diffuse toward defect sites in the graphene lattice, where they form hydrocarbon bonds and become immobilized. This catalytic pathway provides a physically plausible explanation for the substantial enhancement of hydrogen adsorption observed exclusively in Pt-decorated graphene samples.

Quantitative ERDA analysis yields a hydrogen areal density of approximately 10.3E15 protons/cm² for the Pt-decorated graphene after hydrogen exposure. Figure 5(b) shows the corresponding SIMNRA simulation result. When this value is converted into an equivalent hydrogen weight percentage using a monolayer graphene model as a reference, an apparent hydrogen storage capacity of approximately 18.5 wt% is obtained[23, 24]. It is therefore emphasized that the reported hydrogen weight percentage represents an upper-bound effective value integrated over the ERDA probing depth and does not reflect the intrinsic hydrogen uptake of the graphene sheet alone.

In this context, although the derived hydrogen content exceeds commonly cited benchmark values such as the U.S. Department of Energy (DOE) target of 6.5 wt%[25], this discrepancy should not be interpreted as an indication of exceptionally high intrinsic hydrogen storage in graphene. Rather, it highlights the inherent limitations of depth-integrated ERDA measurements when applied to atomically thin materials. Given that the depth resolution of the present ERDA setup exceeds 10 nm, while monolayer graphene is only one atomic layer thick, the detected hydrogen signal inevitably includes contributions from regions beneath the graphene layer.

A more physically reasonable interpretation is that a significant fraction of the hydrogen generated via Pt-catalyzed dissociation and subsequent spillover is not confined to the graphene lattice but diffuses into and becomes trapped within the near-surface region of the underlying $SiO_2$ layer[8, 22]. Consequently, the measured hydrogen signal should be regarded as originating from the combined graphene/$SiO_2$ near-surface system rather than from graphene alone. This interpretation reconciles the experimentally observed hydrogen content with theoretical expectations and underscores the critical roles of Pt catalysis and substrate effects in the hydrogen adsorption mechanism.

Conclusion

In conclusion, ERDA measurements clearly show that Pt-decorated graphene exhibits a measurable hydrogen uptake, whereas graphene subjected only to carbon-ion irradiation displays a negligible hydrogen absorption capability. After hydrogen exposure at a pressure of 14 psi for 5 minutes, the Pt-decorated graphene reaches a saturated hydrogen signal, and no further increase is observed upon subsequent hydrogen exposure cycles. This saturation behavior indicates a well-defined maximum hydrogen uptake under the present experimental conditions.

By contrast, carbon-ion–damaged graphene without Pt decoration does not show a significant increase in hydrogen concentration compared to pristine graphene, indicating that lattice damage induced by MeV carbon ions alone is insufficient to promote hydrogen absorption detectable by ERDA. These results demonstrate that the presence of Pt is a critical factor governing hydrogen uptake in the graphene-based system.

From an instrumentation and measurement perspective, the observed hydrogen signal should be interpreted as an effective, depth-integrated quantity within the ERDA probing depth. The results emphasize the importance of catalytic decoration and substrate effects when evaluating hydrogen-related signals in atomically thin materials using ion-beam analysis techniques.

Figure 1.

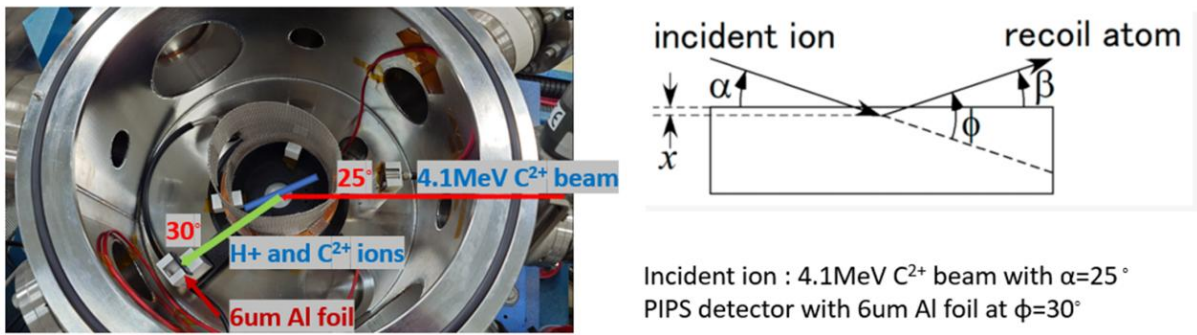

Incident ion : 4.1MeV $C^{2+}$ beam with α=25°
PIPS detector with 6um Al foil at φ=30°

Figure 2.

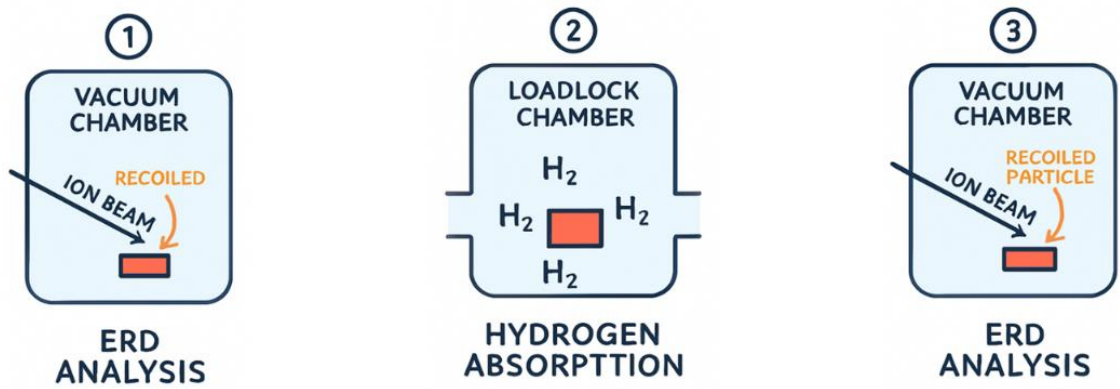

Figure 3.

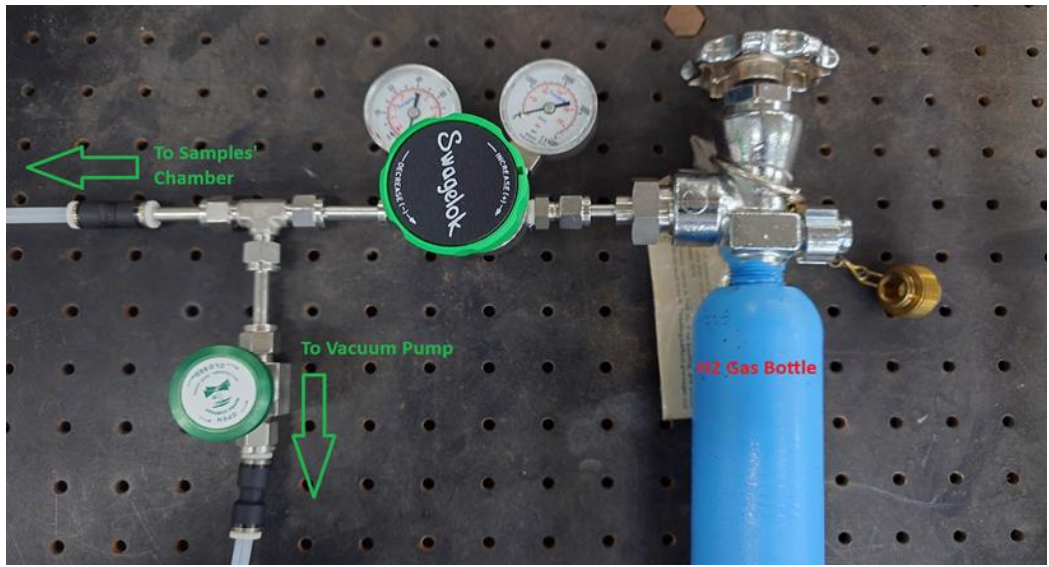

Figure 4.

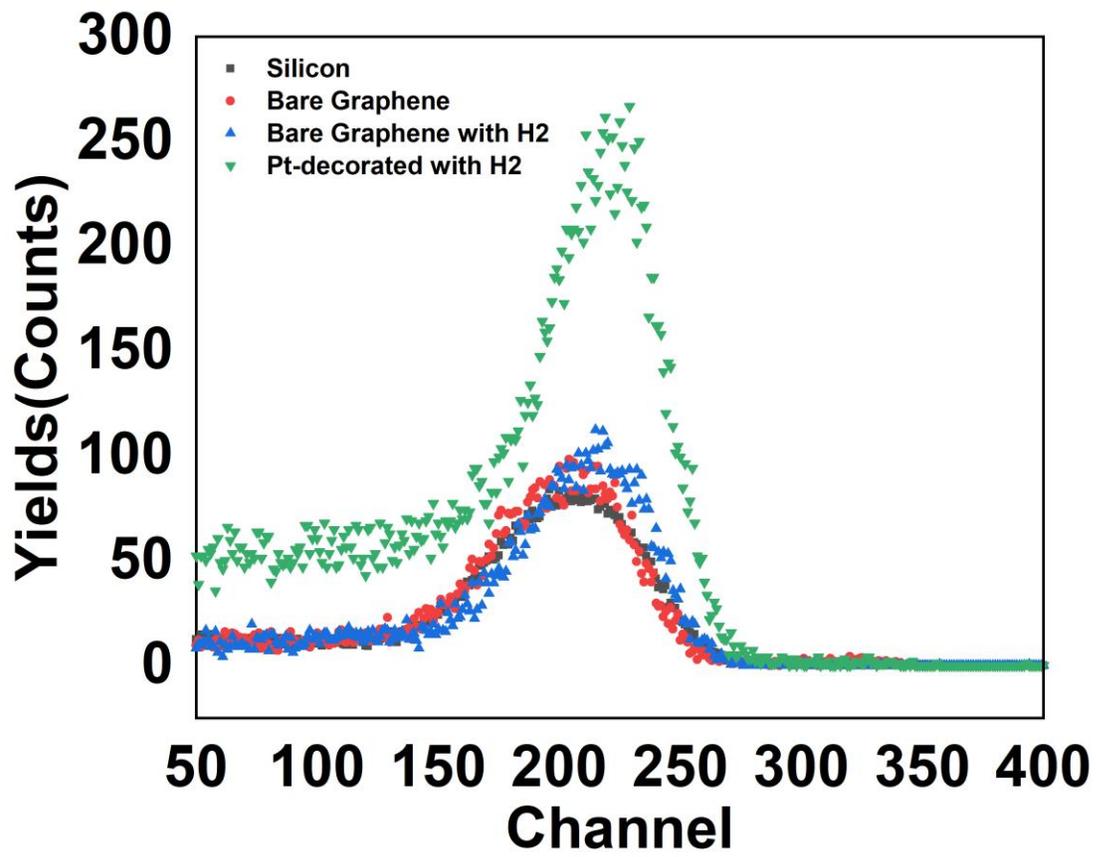

Figure 5a

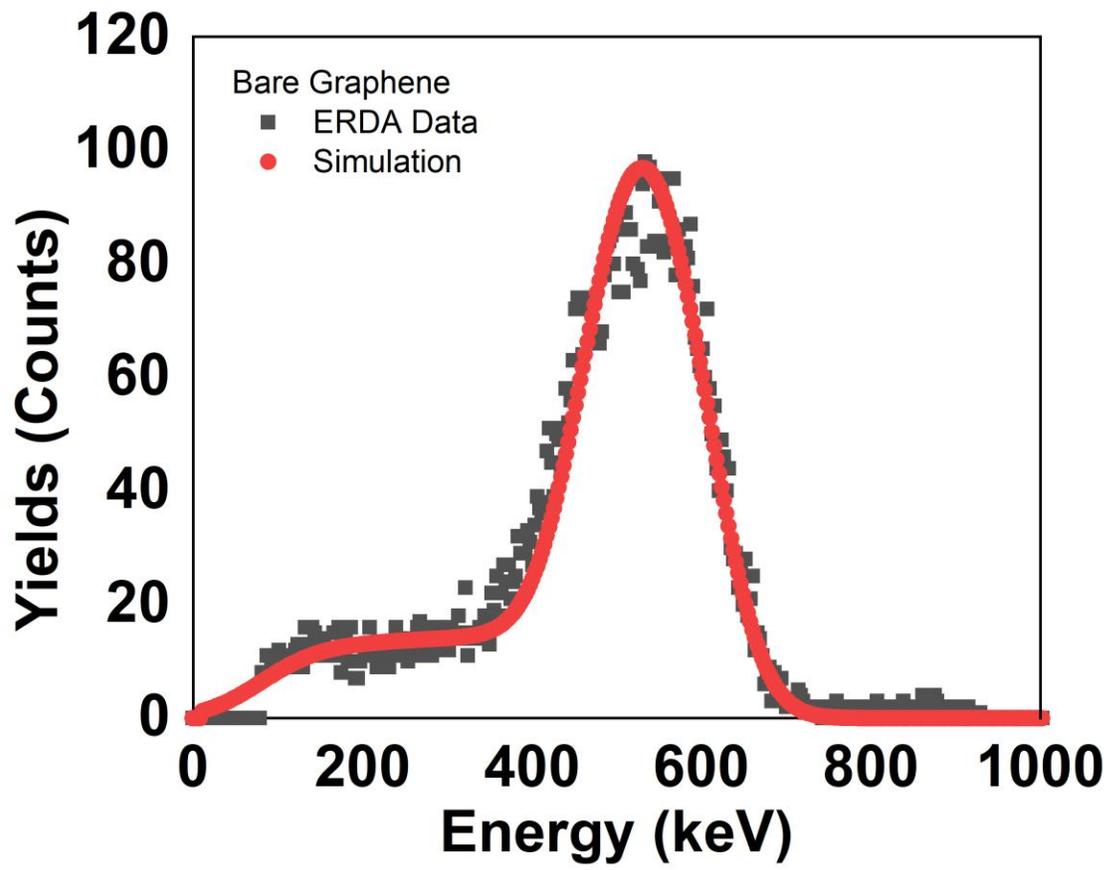

Figure 5b

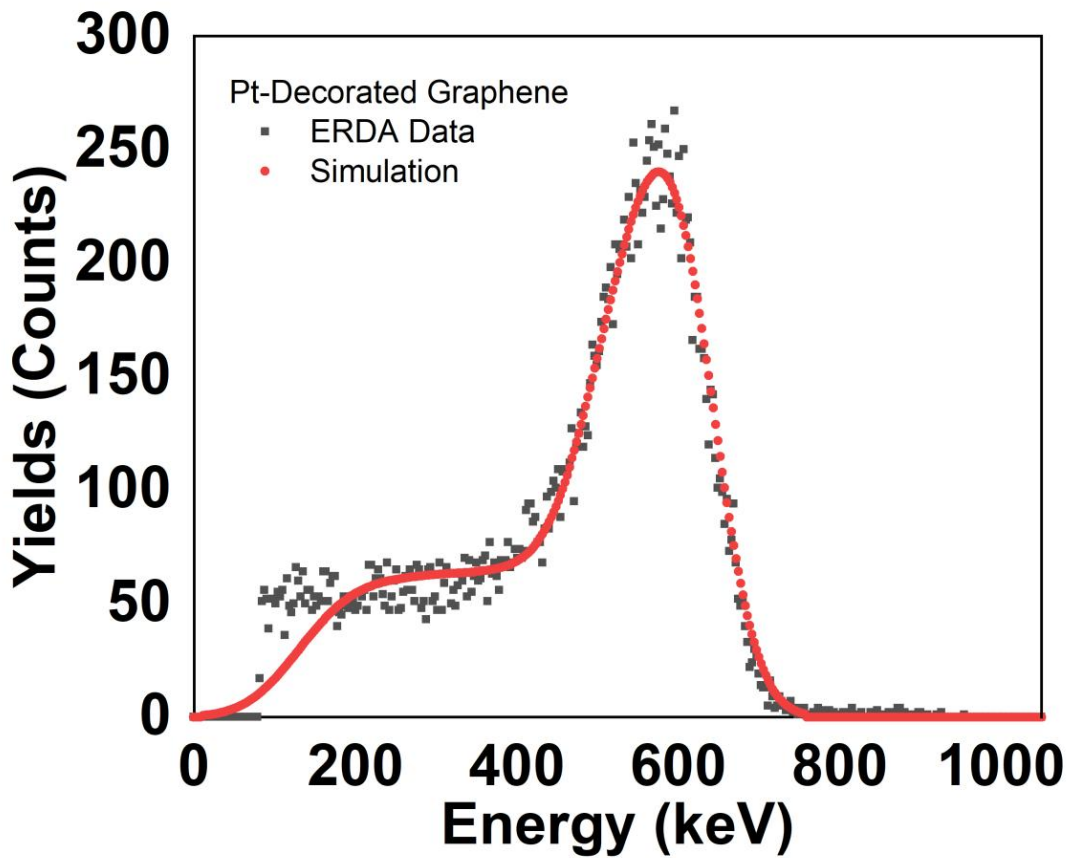

Figure caption

**Figure 1.** Schematic of the ERDA geometry used in this work: a 4.1 MeV $C^{2+}$ beam impinges the sample at a 25° incidence; recoiled H is detected by a PIPS detector placed at a 30° scattering angle and shielded by a 6 μm Al foil to block forward-scattered C ions while transmitting protons.
**Figure 2.** Experimental sequence for hydrogen uptake measurement: (1) first ERD in vacuum with the 4.1 MeV $C^{2+}$ beam; (2) hydrogen exposure in the load-lock at 14 psi $H_2$ for 5 min; (3) evacuation and a second ERD to quantify changes in hydrogen content.
**Figure 3.** Photograph/schematic of the load-lock exposure setup showing the sample transfer, $H_2$ gas fill to 14 psi, and subsequent evacuation before returning the sample to the analysis chamber.
**Figure 4.** ERDA hydrogen recoil spectra comparing (i) pristine Si(001) wafer, (ii) bare graphene (before/after ERDA and after 14 psi $H_2$ exposure), and (iii) Pt-decorated graphene: negligible H signal for Si and bare graphene, and a pronounced increase for Pt/graphene with successive $H_2$ exposures until saturation.

**Figure 5.** SIMNRA-assisted analysis of hydrogen areal density: (a) bare graphene after 14 psi $H_2$ exposure, yielding ~$0.3 \times 10^{15}$ protons cm$^{-2}$ comparable to the Si reference[21]; (b) Pt-decorated graphene after $H_2$ exposure, yielding ~$10.3 \times 10^{15}$ protons cm$^{-2}$ and evidencing Pt-catalyzed uptake via spillover.

Acknowledge

This work was supported by the Ministry of Science and Technology, Taiwan under grants no. NSCT-114-NU-E-007-003-NU and 113-2221-E-007-126

---

i